\newif\ifpreprint

\preprintfalse 

\ifpreprint
\documentclass[journal=jctcce,manuscript=letter]{achemso}
\else
\documentclass[journal=jctcce,manuscript=letter,layout=twocolumn,draft=true]{achemso}
\fi

\usepackage[T1]{fontenc} 
\usepackage{amsmath}
\usepackage{newtxtext,newtxmath}
\usepackage{graphicx}
\usepackage{dcolumn}
\usepackage{braket}
\usepackage{multirow}
\usepackage{threeparttable}
\usepackage{xspace}
\usepackage{verbatim}
\usepackage[version=4]{mhchem} 
\usepackage{comment}
\usepackage{siunitx}
\usepackage{physics}
\usepackage{array}
\usepackage{subcaption}
\usepackage[section]{placeins}
\usepackage{makecell}
\usepackage{mathtools}
\usepackage[dvipsnames]{xcolor}
\usepackage{xspace}
\usepackage{ifthen}
\usepackage{qcircuit}
\usepackage{graphicx,longtable,dcolumn,mhchem}
\usepackage{rotating,color}
\usepackage{lscape}
\usepackage{amsmath}
\usepackage{dsfont}
\usepackage{color,soul}
\newcolumntype{d}{D{.}{.}{-1}}
\usepackage{tabularx}

\newif\ifnohighlight
\nohighlightfalse
\ifnohighlight

\fi

\usepackage[normalem]{ulem}
\usepackage[colorlinks = true,
linkcolor = blue,
urlcolor  = blue,
citecolor = blue,
anchorcolor = blue]{hyperref}
\definecolor{goodorange}{RGB}{225,125,0}
\definecolor{goodgreen}{RGB}{5,130,5}
\definecolor{goodred}{RGB}{220,50,25}
\definecolor{goodblue}{RGB}{30,144,255}

\newcommand{\note}[2]{
	\ifthenelse{\equal{#1}{F}}{
		\colorbox{goodorange}{\textcolor{white}{\footnotesize \fontfamily{phv}\selectfont #1}}
		\textcolor{goodorange}{{\footnotesize \fontfamily{phv}\selectfont #2}}\xspace
	}{}
	\ifthenelse{\equal{#1}{R}}{
		\colorbox{goodred}{\textcolor{white}{\footnotesize \fontfamily{phv}\selectfont #1}}
		\textcolor{goodred}{{\footnotesize \fontfamily{phv}\selectfont #2}}\xspace
	}{}
	\ifthenelse{\equal{#1}{N}}{
		\colorbox{goodgreen}{\textcolor{white}{\footnotesize \fontfamily{phv}\selectfont #1}}
		\textcolor{goodgreen}{{\footnotesize \fontfamily{phv}\selectfont #2}}\xspace
	}{}
	\ifthenelse{\equal{#1}{M}}{
		\colorbox{goodblue}{\textcolor{white}{\footnotesize \fontfamily{phv}\selectfont #1}}
		\textcolor{goodblue}{{\footnotesize \fontfamily{phv}\selectfont #2}}\xspace
	}{}
}

\usepackage{titlesec}

\usepackage[fontsize=11pt]{scrextend}
\titleformat{\section}
{\normalfont\sffamily\bfseries\color{Blue}}
{\thesection.}{0.25em}{\uppercase}

\titleformat{\subsection}
{\normalfont\sffamily\bfseries}
{\thesubsection}{0.25em}{}

\titleformat{\subsubsection}
{\normalfont\sffamily}
{\thesubsubsection}{0.25em}{}

\titleformat{\suppinfo}
{\normalfont\sffamily\bfseries}
{\thesubsection}{0.25em}{}

\titlespacing*{\section}{0pt}{0.5\baselineskip}{0.01\baselineskip}
\titlespacing*{\subsection}{0pt}{0.125\baselineskip}{0.01\baselineskip}
\titlespacing*{\subsubsection}{0pt}{0.125\baselineskip}{0.01\baselineskip}

\newcolumntype{P}[1]{>{\centering\arraybackslash}p{#1}}

\newcommand{\CEISAM}{Nantes Universit\'e, CNRS,  CEISAM UMR 6230, F-44000 Nantes, France}

\author{Vincent Delmas} 
\affiliation[UN, Nantes]{\CEISAM}    
\author{Alessandro Nicola Nardi}
\affiliation[UN, Nantes]{\CEISAM}    
\author{Isabella C. D. Merritt}
\affiliation[UN, Nantes]{\CEISAM}
\alsoaffiliation[UB, Bordeaux]{Univ. Bordeaux, CNRS, Bordeaux INP, Institut des Sciences Moléculaires (ISM), F-33405, Talence, France}        
\author{Anthony Ferté}
\affiliation[UN, Nantes]{\CEISAM}    
\author{Ignacio Fdez. Galván} 
\affiliation[Uppsala]{Department of Chemistry – BMC, Uppsala University, P.O. Box 576, SE-751 23 Uppsala, Sweden}
\alsoaffiliation[Uppsala2]{Department of Chemistry – Ångström, Uppsala University, P.O. Box 523, SE-751 20 Uppsala, Sweden}
\author{Morgane Vacher} 
\affiliation[UN, Nantes]{\CEISAM}    
\email{morgane.vacher@univ-nantes.fr}

\setkeys{acs}{usetitle=true}

\let\oldmaketitle\maketitle
\let\maketitle\relax
\title[Reduced Dimensionality Dynamics]{Automated selection of nuclear coordinates for reduced dimensionality non-adiabatic dynamics}
\date{\today}

\keywords{mots-clefs}

\begin{tocentry}
	\centering
	\includegraphics[width=\textwidth,draft=false]{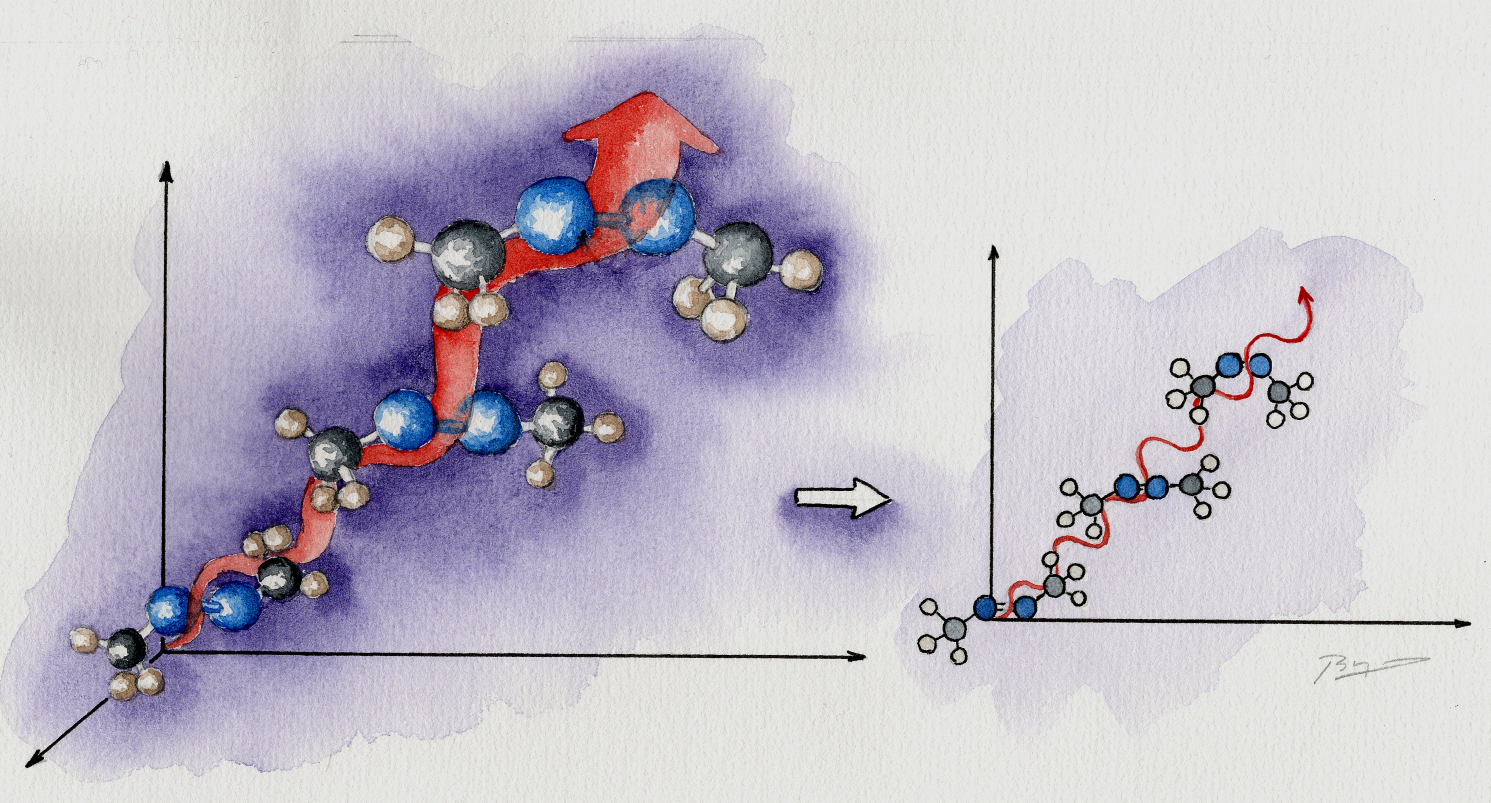} 
\end{tocentry}

\begin{document}

	\ifpreprint
	\else
	\twocolumn[
	\begin{@twocolumnfalse}
		\fi
		\oldmaketitle
		
		\begin{abstract}

		\noindent Poor scaling of dynamics simulations with number of dimensions is currently a major limiting factor in the simulation of photochemical processes. In this work we investigate ways to reduce the dimensionality of many-atom systems with a view towards enhancing computational efficiency while maintaining accuracy. Using mixed quantum-classical Trajectory Surface Hopping (TSH) simulations of three photo-reactive molecules - \textit{trans}-azomethane (tAZM), butyrolactone (Bulac), and furanone (Fur) - we explore two different dimensionality reduction techniques: Principal Component Analysis (PCA) and Normal Mode Variance (NMV). Dynamics simulations are run in full dimensionality and reduced dimensionality, employing either PCA or NMV, and the impact of the dimensionality reduction on selected electronic and geometric properties of the dynamics is evaluated. For all three molecules, both PCA and NMV can be used to select lower-dimensional spaces in which the full-dimensionality dynamics results are reproduced. PCA reduction outperforms NMV in all systems, allowing for a more significant dimensionality reduction without loss of accuracy. The improved accuracy of PCA is, for tAZM, mostly seen in the electronic properties while for both Fur and Bulac the advantage is clear in the ring-opening reaction itself as well. The present approach opens routes to simulation of larger photochemically relevant systems, through the use of automated dimensionality reduction, avoiding human bias.
		\end{abstract}
		
		\ifpreprint
		\else
	\end{@twocolumnfalse}
	]
	\fi
	
	\ifpreprint
	\else
	\small
	\fi
	
	\noindent
	
	\section{Introduction}
	\label{sec:Intro}
    
        \noindent Modelling excited-state dynamics allows us to understand how fundamental processes such as photochemical reactions, energy transfer, and transitions between electronic states occur. This is key for the understanding of natural processes and development of new technologies based on light-absorption, a field of significant current interest e.g. for green synthesis and opto-electronic devices \cite{Balzani2008,Oelgemller2016}. Advances in computational chemistry have enabled detailed exploration of small molecular systems with remarkable precision \cite{Pathak2020,Mai2020,Latka2019}. However, accurate modelling of excited-state dynamics processes often proves challenging due to the inherent increase in computational cost as systems increase in size, with the dimensionality of the conformation space for a (non-linear) molecule being $D = 3N_{at}-6$, where $N_{at}$ is the number of atoms. The cost of dynamics simulations generally scales very poorly with dimensionality, with the severity of the scaling depending on the method used: in grid-based molecular quantum dynamics simulations, exponential scaling on $D$ is often labelled as ``the curse of dimensionality" \cite{Zauleck2016,Liu2021}. Practically, this limits full-dimensional dynamics studies to small molecules, with a different practical upper limit for $N_{at}$ depending on the excited state dynamics method employed along with any additional approximations within. 
	
        \noindent One approach to address the limitations imposed by increasing molecular size, often employed for fully quantum excited state dynamics methods such as Multi-Configuration Time-Dependent Hartree (MCTDH) \cite{meyer1990multi}, is to instead run the dynamics on a model of the molecule consisting of only a selected subset of dimensions, expected to be representative of the essential features of the dynamics. Determining the optimal subset of dimensions for such simulations is a critical aspect of model construction in MCTDH and other dynamics methodologies, with a compromise being necessary between accuracy and cost. Chemical knowledge of the system and process to be modelled, along with symmetry considerations \cite{Worth2008}, often guides the selection of relevant degrees of freedom to be included in the model. Unfortunately, this lack of a systematic and impartial approach implies a natural bias in the dynamics simulated, potentially limiting the accuracy of such models.
	
        \noindent Prior literature has explored the idea of identifying the most important nuclear dimensions in a systematic manner, with the aid of lower-level dynamics such as the mixed-quantum-classical trajectory surface hopping (TSH) method \cite{Mai2019}. For instance, G\'omez \textit{et} al. \cite{Gmez2019} introduced an iterative approach using a feedback loop between MCTDH and linear vibronic coupling TSH \cite{Plasser2019}, to identify the minimum number of nuclear dimensions and electronic states to produce accurate dynamics. In their study, the least important nuclear modes are identified according to the amplitudes of key associated parameters (namely, the inter and intrastate coupling to other modes). However, they note that this selection process could be carried out in a number of alternative ways, such as by using excited-state properties in their model, or identification of unimportant modes from the dynamics simulations. 
	
        \noindent In the present work, we evaluate the use of two different post-simulation analysis methods, namely, Principal Component Analysis (PCA) \cite{Pearson1901, wold1987principal} and Normal Mode Variance (NMV), to automatically and systematically reduce dimensionality. 
        PCA and other linear and non-linear methods have already been used quite extensively \cite{Zauleck2016, Capano2017, Tavadze2017, Zhu2022, Gmez2019, Li2017, Li2018} in the field of molecular dynamics. However, this has been primarily with the aim of understanding multidimensional data created by dynamics simulations - i.e., for post-processing of high-dimensional data.
 In contrast, we propose to use PCA and NMV to extract meaningful information from dynamics performed in full dimensionality using for instance the mixed quantum-classical TSH method \cite{Tully1990,Tully1971} and to identify the dimensions which are the most important when describing the reaction, based on this analysis.  Selecting a relevant reduced dimensional space through such a protocol could be useful when aiming to run future simulations modelling photochemical reactions with higher accuracy (but higher cost) methods, for example using an improved electronic structure level, or in fully quantum dynamics simulations.
  Importantly, the suggested protocol assumes that the lower-level dynamics is able to capture the relevant features of the investigated process. Previous theoretical studies have benchmarked the surface hopping method against more accurate dynamics methods and demonstrated its suitability for simulating photochemical reactions \cite{Gomez-2024,Ibele-2020,Janos-2023}, although it may be less suited for complex atto-photochemical reactions \cite{Tran-2024}. In the present work, a given combination of dynamics method (TSH) and electronic structure level (see Section 2.3) is used throughout. The dynamics simulated in the reduced dimensional space is compared to the full-dimensional results, to evaluate the quality of PCA/NMV as a dimensionality reduction technique.

\begin{figure}[H]
        \centering
        \includegraphics[width=0.9\linewidth]{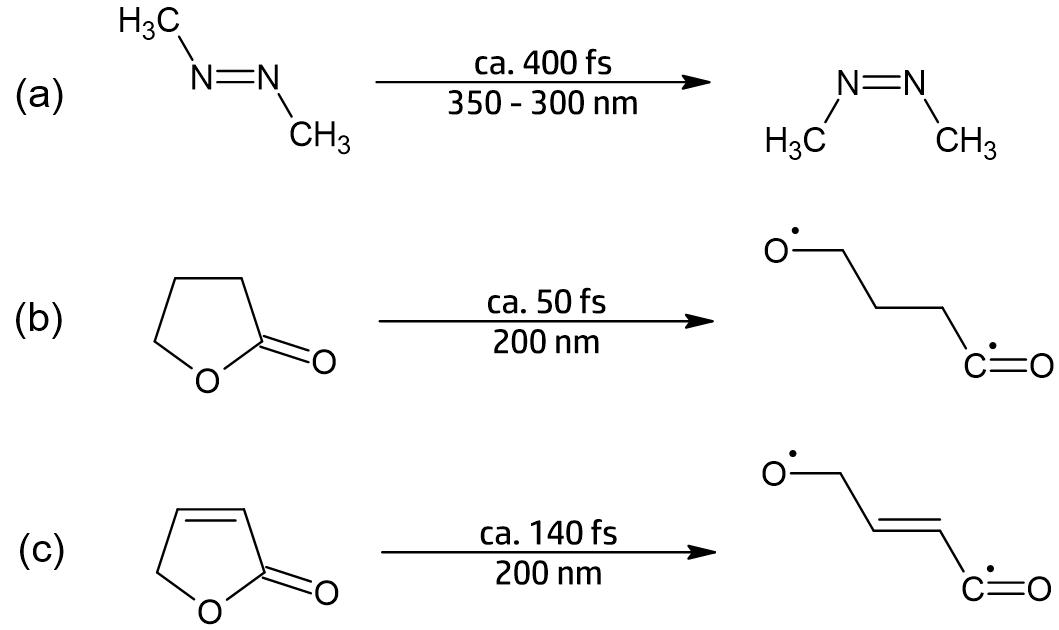}
        \caption{(a) \textit{trans}-to-\textit{cis} isomerisation of azomethane, (b) ring opening of butyrolactone, and (c) ring opening of furanone. Reaction times and excitation wavelength are taken from refs \citenum{ruckenbauer2010} and \citenum{Schalk2020}.}
        \label{fig:reactions} 
    \end{figure}

    \noindent The dimensionality reduction protocol is tested on the three photochemical reactions illustrated in Figure \ref{fig:reactions}, selected to cover a range of photochemical situations. The first reaction (Figure \ref{fig:reactions}(a)) is the \textit{trans}-to-\textit{cis} photoisomerisation of azomethane (tAZM), a prototypical double bond rotational isomerisation which occurs on a 400 fs timescale according to prior simulations.\cite{ruckenbauer2010} The second (Figure \ref{fig:reactions}(b)) and third (Figure \ref{fig:reactions}(c)) reactions are examples of bond breaking occuring during the ring-opening reactions of butyrolactone (Bulac) and furanone (Fur) respectively after excitation to the second excited state. These reactions differ in the mechanism of ring-opening, Bulac undergoing rapid ballistic bond dissociation within \hbox{50 fs}, while bond dissociation in Fur occurs more slowly and gradually, \textit{ca.}~\hbox{140 fs}, due to competition with ring-puckering.\cite{Schalk2020} 
    
    \vspace{-5pt}

    \noindent This article is structured as follows: in Section 2 the methodology is introduced, including the dimensionality reduction procedure, the electronic and geometric properties used to evaluate the reactions, as well as the computational details. Following this, Section 3 presents the findings on the three studied molecules. Subsequently, Section 4 delves into the analysis, before the final conclusions are given in Section 5.

	\section{Methodology}
	\label{sec:methodo}

    \subsection{Dimensionality reduction procedure}
	
\begin{figure}
		\centering
		\includegraphics[width=0.9\linewidth]{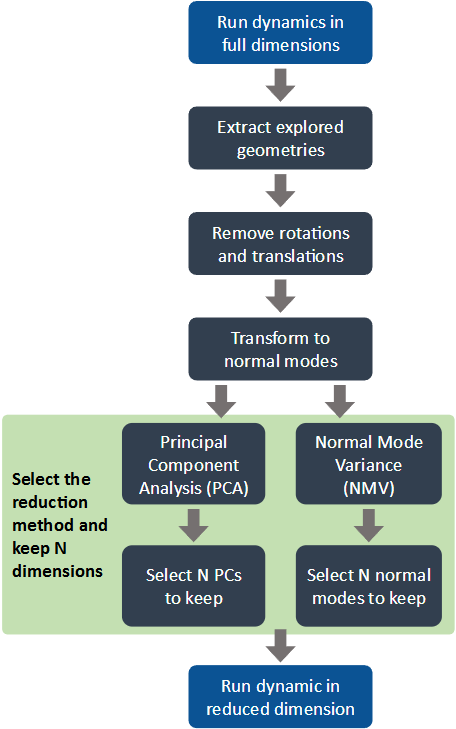}
		\caption{Workflow of the dimensionality reduction procedure, starting from full dimensionality simulations.}  
		\label{fig:workflow}
	\end{figure}

        The overall methodology used in this work is outlined in Figure \ref{fig:workflow}. First the dynamics are performed in full dimensionality and visited geometries along the trajectories are extracted. As mixed-quantum classical simulations are typically performed in Cartesian coordinates, full dimensionality here actually refers to $3N_{at}$ dimensions. The 6 additional dimensions compared to the $3N_{at} - 6$ conformation space of a non-linear molecule arise from rotation and translation - quantum dynamics simulations are typically run in normal modes, without these motions. The Kabsch algorithm \cite{kabsch1988} is then subsequently applied to remove any rotational and translational differences from all geometries compared to the Franck-Condon geometry (assigning different weights to the different atoms depending on their mass). The Cartesian geometries are then converted to normal mode coordinates, using the ground-state minimum as the reference geometry and the normal mode vectors calculated at that reference geometry. One dimensionality reduction method is finally applied: PCA or NMV. The Principal Components (PCs) or the Normal Modes (NMs) with the greatest variance are selected, to form a new basis to simulate the dynamics in reduced dimensions. 
        
\noindent The first dimensionality reduction approach used, PCA, \cite{Pearson1901, wold1987principal} is a linear transformation method, known to be a robust technique commonly used in machine learning, statistics, data analysis \cite{bro2014principal, karamizadeh2013overview}, and in dynamics \cite{Zauleck2016, Capano2017, Tavadze2017, Zhu2022}. Its primary usage is to transform a dataset with a high number of variables into a new set of uncorrelated variables, called principal components (PC), while capturing the maximum variance present in the data. The created lower number of PC retain the maximum amount of variance possible compared to the original dataset, and are ordered by decreasing order of variance: the first PC explains the largest amount of variance in the dataset, followed by the second, and so on. See Supporting Information (SI), Section 1 for more details.

\noindent The second dimensionality reduction approach used, NMV, is a very straightforward technique. It consists of computing the variance of all geometries explored across the full-dimensional dynamics along each normal mode (as calculated at the initial geometry of the molecule). The normal modes are then ordered according to their explained variance, in decreasing order, and those containing the lowest variance values are then consecutively removed to reduce the dimensionality. See SI, Section 2 for more details.

        \noindent The non-adiabatic dynamics is then simulated in the reduced subspace selected by either PCA or NMV, using the same methodology, i.e., TSH, and computational settings as those in full dimensionality. It is important to realise that the electronic structure properties used in TSH, such as the energies and gradients, are not calculated directly in the lower dimensional subspace. Instead, at each time step, the gradient calculated analytically in the full dimensional space is transformed - on the fly - into normal modes and projected onto the reduced subspace (using either the PC basis from PCA or the normal mode basis containing only the most relevant modes as determined by the NMV procedure). After this reduction step, the gradient is transformed back into Cartesian coordinates and used in the propagation of Newton's equation of motion via the velocity-Verlet algorithm. In this fashion, the geometry is naturally propagated within the reduced dimensional subspace. 
        An alternative method to remove discarded dimensions from the dynamics is to project, at each time step, the new geometry predicted using the full-dimensional gradient into the reduced space. Both the reduction on the gradients and the geometries were implemented and tested, leading to the same reduced dimension dynamics. In the present work, the reduction was applied on the gradients (forces). Importantly, initial conditions (geometries and velocities) need to be reduced before starting the dynamics. See SI section 3 for a detailed step-by-step protocol for the dynamics in reduced dimensionality.
        
        \noindent Since the electronic structure calculations are performed in full dimension, there is no increase in speed associated with running TSH in lower dimensions as implemented here; our purpose is solely to provide a proof-of-concept of high-quality dynamics run in the minimal number of dimensions, as well as to compare the two selection techniques (PCA/NMV) for dimensionality reduction protocols.
	
	\subsection{Dynamical Properties}
	
	\begin{table*}[t]
		\centering
		\begin{tabularx}{0.9\linewidth}{lX}
			\hline \hline
			\multicolumn{1}{c|}{\textbf{\:\:\:\:Properties\:\:\:\:}} & \multicolumn{1}{c}{\textbf{\textit{Trans}-azomethane}} \\ \hline
			\multicolumn{1}{c|}{\textbf{Electronic}} &
			\begin{tabular}[c]{@{}l@{}}
				- $S_1$ and $S_0$ state population crossing time (in fs).\\ 
				- $S_0$ state population (in \%) at the end of the simulation (280 fs).
			\end{tabular} \\
			\multicolumn{1}{c|}{\textbf{Geometric}} &
			\begin{tabular}[c]{@{}l@{}}
				- Time (in fs) to reach half of final \textit{cis} yield.\\ 
				- \textit{cis} yield (in \%) at the end of the simulation (280 fs).
			\end{tabular} \\ \hline \hline
			\multicolumn{1}{c|}{\textbf{\:\:\:\:Properties\:\:\:\:}} & \multicolumn{1}{c}{\textbf{Butyrolactone}} \\ \hline
			\multicolumn{1}{c|}{\textbf{Electronic}} &
			\begin{tabular}[c]{@{}l@{}}
				- Time (in fs) for the $S_2$ population to reach the inflection point.\\ 
				- $S_0$ state population (in \%) at the end of the simulation (100 fs).
			\end{tabular} \\
			\multicolumn{1}{c|}{\textbf{Geometric}} &
			\begin{tabular}[c]{@{}l@{}}
				- Time (in fs) to reach half of the final C-O bond break yield.\\
				- C-O bond break yield (in \%) at the end of the simulation (100 fs).
			\end{tabular} \\ \hline \hline
			\multicolumn{1}{c|}{\textbf{\:\:\:\:Properties\:\:\:\:}} & \multicolumn{1}{c}{\textbf{Furanone}} \\ \hline
			\multicolumn{1}{c|}{\textbf{Electronic}} &
			\begin{tabular}[c]{@{}l@{}}
				- Time (in fs) to reach half of the final $S_2$ population yield.\\ 
				- $S_0$ state population (in \%) at the end of the simulation (150 fs).
			\end{tabular} \\
			\multicolumn{1}{c|}{\textbf{Geometric}} &
			\begin{tabular}[c]{@{}l@{}}
   			- Time (in fs) to reach half of the final C-O bond break yield.\\ 
				- C-O bond break yield (in \%) at the end of the simulation (150 fs).
			\end{tabular}
			\\ \hline \hline
		\end{tabularx}
		\caption{Electronic and geometric properties chosen to evaluate the quality of the dimensionality reduction for each molecule.}
		\label{table:descriptors}
	\end{table*}

        \noindent The time evolution of the electronic populations, as well as the time evolution of the key nuclear coordinates, for full- and reduced-dimensionality dynamics obtained with PCA and NMV methods are presented in SI. To quantitatively assess the accuracy of the dimensionality reduction in comparison to the full dimensional ($3N_{at} - 6$) dynamics, two electronic and two geometric properties were selected for each molecule. These properties are presented in Table \ref{table:descriptors}. For each pair of electronic and geometric properties, one is chosen to quantify the timescale of the process, and the other its yield. The accuracy of a dynamics simulation is assessed by calculating the absolute relative error (ARE) for each property $P$:
	
	\begin{equation}
		\text{ARE} = \left| \frac{P_{\text{full}} - P_{\text{red}}}{P_{\text{full}}} \right|
		\label{eq:ARE}
	\end{equation}
     \vspace{5pt}
	
        \noindent It involves comparing the value of the property calculated from the ensemble of trajectories, for both the full $P_{\text{full}}$ and reduced dimensionality $P_{\text{red}}$ dynamics simulations.
	
        \noindent Finally, an overarching error (labelled as the ``total" error) is defined as the average of the ARE over all properties:
	
	\begin{equation}
		\text{ARE}_{\text{total}} = \frac{\text{ARE}_{\text{el1}} + \text{ARE}_{\text{el2}} + \text{ARE}_{\text{geom1}} + \text{ARE}_{\text{geom2}}}{4}
		\label{eq:total}
	\end{equation}
	
        \noindent With \textit{el} and \textit{geom} denoting the electronic and geometric properties, respectively. The values of each error (electronic, geometric and total), along with the average values of each property (ARE) used to evaluate the reactions (Table \ref{table:descriptors}), can be found in SI.

    \subsection{Computational details}
    For each selected set of dimensions for each molecule, 100 trajectory surface hopping \cite{Tully1971,Tully1990} trajectories corresponding to 100 sampled initial conditions were simulated using the OpenMOLCAS code.\cite{OpenMolcas} The initial conditions were generated using SHARC,\cite{SHARC3} sampled from the Wigner distribution without temperature broadening at the Franck-Condon geometry, using harmonic frequencies calculated using OpenMOLCAS (see below for specifics of the electronic structure for each molecule).
    A set of 100 trajectories was found enough to converge the dynamics in full dimensionality, see Ref.~\citenum{Merritt2023} and its SI.
    \textit{Trans}-azomethane was simulated for a total of 280 fs, butyrolactone for a total of 100 fs, and furanone for a total of 150 fs.  A timestep of 20 a.u. (ca. 0.48 fs) for the integration of Newton's equations of motion through the velocity-Verlet algorithm was used, with the integration of the electronic wavefunction at each time step split into 96 substeps. The energy based Persico-Granucci decoherence correction, with a decay factor of 0.1 was used throughout.\cite{Grannucci2007} Energy conservation has been checked for all simulations, checking for active space stability, intruder states, and other simulation artifacts. A maximum of 10\% of the trajectories were allowed to be rejected before an ensemble was considered to be no longer representative. Importantly, the dimensionality reduction does not lead to an increase in the number of trajectories not conserving total energy in comparison to simulation in full dimensionality.

    \noindent The location of hops was determined using the Fewest-Switches Surface Hopping algorithm\cite{Tully1990} and after hopping energy conservation was ensured by re-scaling the kinetic energy along the velocity direction. The Hammes-Schiffer-Tully scheme\cite{HammesSchiffer1994} was used to approximate the non-adiabatic couplings at each timestep, using the biorthonormal wavefunction overlap scheme as implemented in OpenMOLCAS.\cite{Merritt2023, OpenMolcas}
	
    \noindent For each molecule, electronic structure calculations were carried out using the state-averaged complete active space self-consistent field (SA-CASSCF) method, with active spaces and state-averages defined as follows: SA2(6e, 4o) for \textit{trans}-azomethane, SA3(12e, 10o) for furanone, and SA3(10e, 8o) for butyrolactone. The active spaces are identical to those selected in Refs \citenum{Schalk2020} and \citenum{Merritt2023}, with full orbital descriptions found in the SI of Ref. \citenum{Merritt2023}. For all molecules the 6-31G(d) basis set\cite{hehre1972} with a single set of \emph{3d} polarization functions\cite{hariharan1973} on the C and N centers was used. Resolution-of-identity based on the Cholesky decomposition was used throughout to speed up electronic structure calculations.\cite{Aquilante2007}

    \section{Results}
    \label{sec:results}
    
    \subsection{\textit{Trans} $\rightarrow$ \textit{cis} isomerisation of azomethane (tAZM)}

    \noindent The first photochemical reaction studied is the \textit{trans}-to-\textit{cis} photoisomerization of azomethane, which occurs after excitation to the first excited state. According to prior simulations, \cite{sellner2010, ruckenbauer2010} this photochemical transformation takes place over a period of ca.~400 fs, through a rotation about the central \ce{C-N=N-C} moiety. The excited state decay takes place through a conical intersection located roughly halfway between the two isomers, with a central dihedral around 90°.
    
            \begin{figure*}[t]
		\centering
		\includegraphics[width=\linewidth]{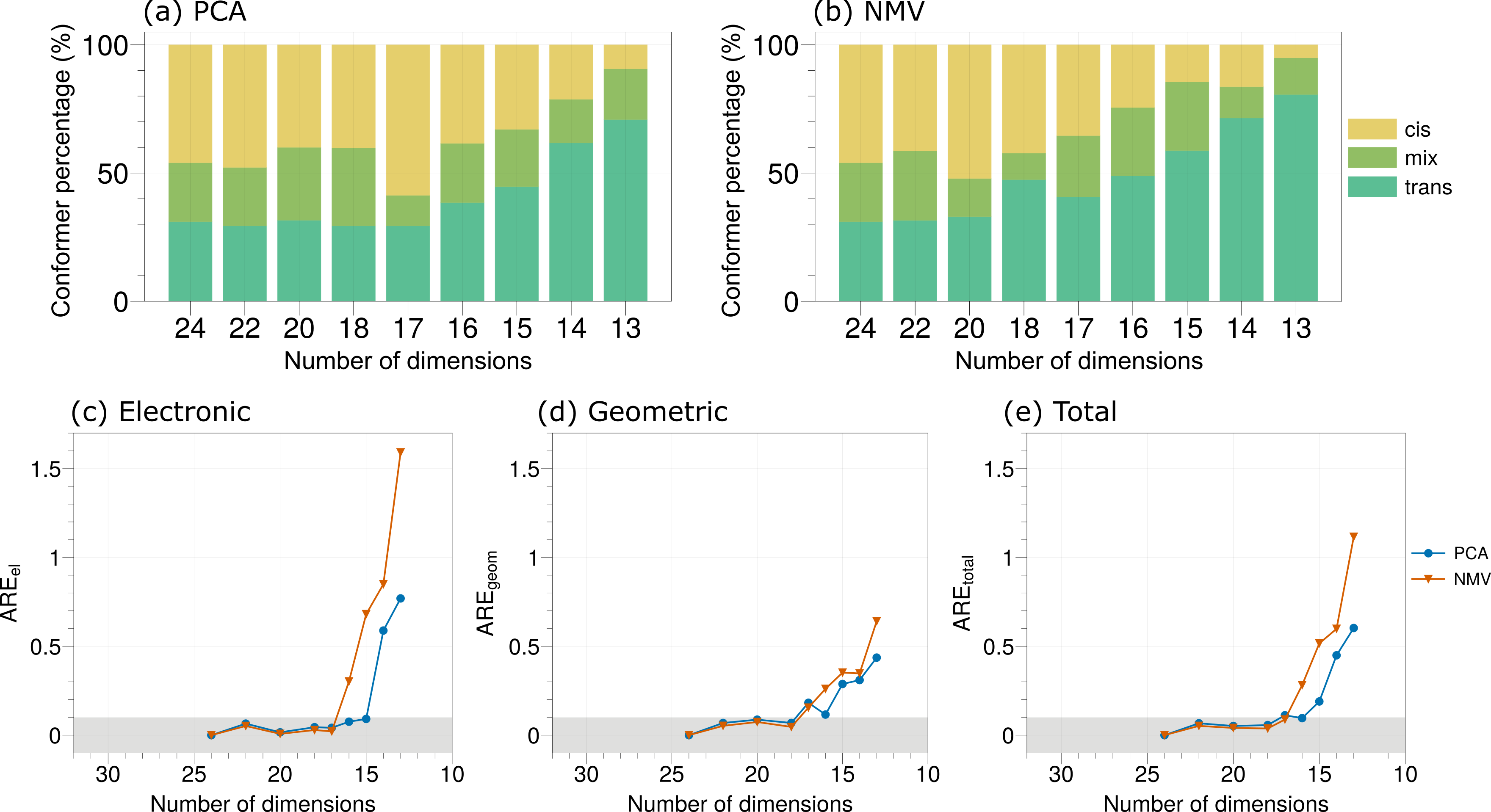}
            \caption{Simulation results for \textit{trans}-azomethane. Distribution (\%) of the \textit{trans}, \textit{mix}, and \textit{cis} conformers at the end of the dynamics simulations as a function of number of dimensions, for both PCA (a) and NMV (b) reduction methods. Evolution of the electronic (c), geometric (d) and the total mean (e) errors, as a function of the number of dimensions, for each reduction method. In (c), (d), and (e), the shaded area represents 10\% of relative error compared to the dynamics performed in full dimensionality.}
		\label{fig:pca_nmv_tazm}
	\end{figure*}
 
    \noindent The time evolutions of the electronic populations of $S_0$ and $S_1$ states, as well as the time evolutions of the dihedral torsion angle \ce{C-N=N-C}, which describes the reaction coordinate, for the full- and reduced-dimension dynamics are presented in SI. The compositions of the principal components in terms of normal modes, as well as graphical representations of the first four components calculated from the full-dimensional dynamics are reported in the SI. The first components consist of CH$_3$ rocking, NN stretching, CNN asymmetric bending and NN torsional motions.
    
\noindent In order to follow the progress of the reaction, the trajectories are classified into one of three conformers based on the value of this central dihedral angle: \textit{cis} for angles under 45°, \textit{mix} for those between 45° and 135°, and \textit{trans} for angles of 135° or more. The fractions of conformers at the end of the simulations using this classification are shown in Figure \ref{fig:pca_nmv_tazm} for both the (a) PCA and (b) NMV dimensionality reduction methods. A finer classification of the geometries based on the central dihedral angle leads to the same conclusions (see SI).

    \noindent Both the NMV and PCA methods allow for some level of dimensionality reduction to be carried out without significant changes to the conformer distributions. In general, the distributions of \textit{trans}, \textit{mix}, \textit{cis} at the end of the dynamics, shown in Figure \ref{fig:pca_nmv_tazm}(a) and (b), evolve quite similarly between the two reduction methods as the number of dimensions decreases. Differences to the reference final distribution obtained in full dimensionality ($3N_{at} -6 = 24$) begin to be seen using fewer than 20 dimensions, suggesting reduction from 24 to 20 dimensions is feasible. With 16 dimensions and fewer, the final balance of isomers shifts more towards the \textit{trans} isomer: i.e., the \textit{trans}-to-\textit{cis} isomerisation is reduced. While from these conformer distributions it seems like the PCA method may allow for slightly more dimensionality reduction to 18 dimensions, it is hard to conclude which method allows for more accurate dimensionality reduction based on this property analysis alone.

    \noindent Figure \ref{fig:pca_nmv_tazm}(d) shows the quantitative error obtained for the two chosen geometrical properties. We define accuracy in the dynamics results to be when ARE (compared to the full dimensional dynamics) is below a threshold of 10\%.
    This threshold is significantly larger than the variance expected from not fully converged simulations (see Ref.~\citenum{Merritt2023} and its SI).
     We see for the NMV a loss of accuracy at 18 dimensions. The PCA result is less clear, with an initial loss of accuracy at 17 dimensions followed by a decrease in ARE with 16 dimensions.
    
    \noindent Figure \ref{fig:pca_nmv_tazm}(c) shows the absolute relative error in  the electronic properties. Both reduction methods show a clear lowest dimensionality point beyond which the ARE dramatically increases. The PCA method maintains accuracy when decreasing to 15 dimensions, beyond which the ARE significantly increases, while the NMV method remains accurate only up to 17 dimensions. We note that the AREs are, in general for \textit{trans}-AZM, lower for the geometric properties compared to the electronic ones.
    
        \begin{figure}
    \centering
    \includegraphics[width=0.9\linewidth]{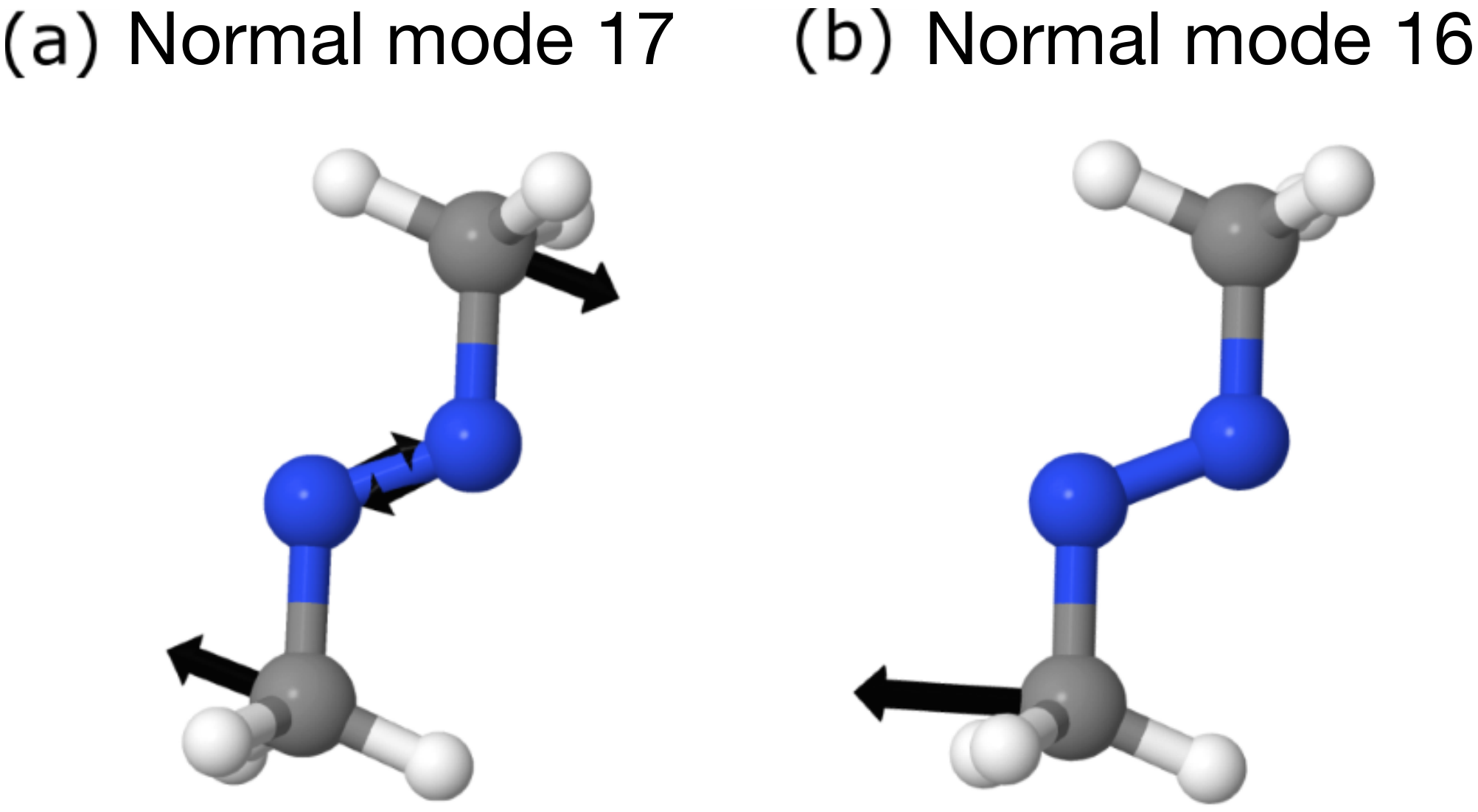}
        \caption{The two normal modes (mass weighted, hydrogen contributions hidden for ease of interpretation) removed when decreasing from 17 to 16 (a) and 16 to 15 (b) dimensions for the azomethane molecule using the NMV method. The normal modes are numbered according to their explained variance (in decreasing order).}
    \label{fig:nm_azm}
    \end{figure}

    \noindent For the NMV method, the total error dramatically increases when decreasing from 17 to 15 dimensions. The normal modes (mass weighted, with contributions on hydrogens hidden) removed for this dimensionality reduction are illustrated in Figure \ref{fig:nm_azm}. It is noted that the normal modes are numbered according to their explained variance in decreasing order, consistently with the PCA modes. The first mode (a) appears to represent a symmetrical \ce{C-N=N} angle opening and reduction of the N=N bond length, while the second mode (b) is less straightforward, possibly indicating the opening of a single C-N=N angle. These two normal modes are likely to play significant roles in the \textit{trans}-to-\textit{cis} isomerisation process. However, because the correlation between their variance and the main reaction pathway is not direct, these were removed, showing the limit of an approach only based on the variance. This is likely to be a notable problem for motions which vary only a small amount within the dynamics, but which are still important for the photochemical mechanism: previously suggested to be the case for the length of the N=N bond.\cite{sellner2010} Furthermore, it should be noted that the significance of all the preceding normal modes removed must also be considered, and the removal of these two modes may simply mark a pivotal moment where the dynamics undergoes a drastic change due to the overall set of modes removed.
        
    \noindent Clearly, the total error as shown in Figure \ref{fig:pca_nmv_tazm}(e) is more influenced by the electronic errors and unambiguously shows that the PCA method exhibits a greater capacity to operate at lower dimensions when compared to the NMV method. Based on this total error, we predict that it would be possible to reduce the dimensionality required to describe this \textit{trans}-to-\textit{cis} photoisomerization of azomethane from 24 ($3N_{at}-6$) to only 16 (18) dimensions, using the PCA (NMV) reduction technique. 

	\subsection{Ring opening of butyrolactone (Bulac)}
 
     \begin{figure*}[t]
    \centering
    \includegraphics[width=\linewidth]{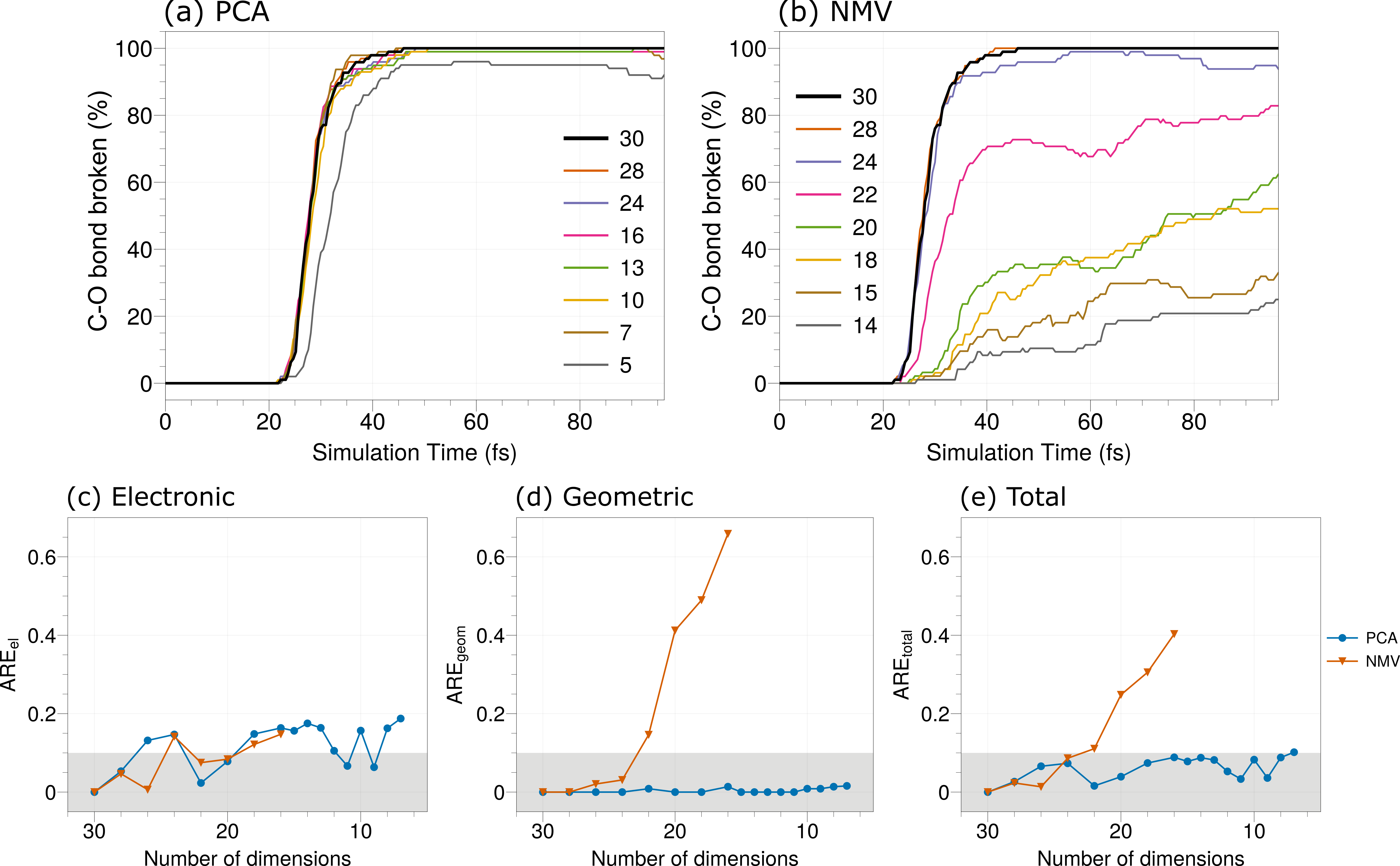}
        \caption{Simulation results for butyrolactone. Top panel: evolution (\%) of C-O bond break as a function of time (fs) for each dimensionality, using the PCA (a) and the NMV (b) method. Evolution of the electronic  (c), geometric  (d) and  total mean errors (e), as a function of the number of dimensions for both applied reduction methods. In (c), (d), and (e), the shaded area represents 10\% of relative error compared to the dynamics in full dimensions.}
    \label{fig:pca_nmv_blac}
    \end{figure*}
    
    	     \begin{figure}
		\centering
		\includegraphics[width=0.9\linewidth]{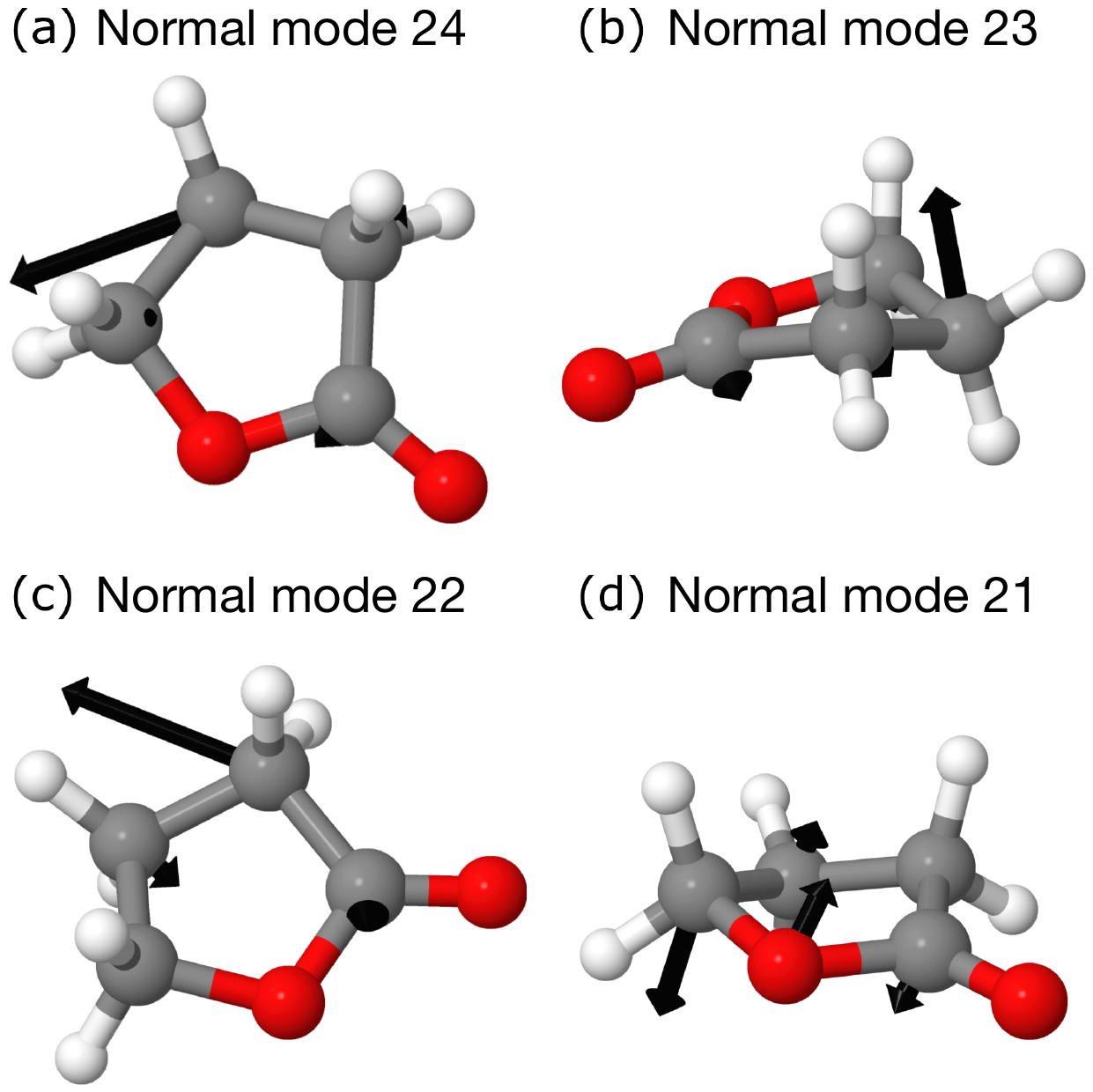}
		\caption{The four normal modes removed when decreasing from 24 to 22 dimensions (a, and b) and from 22 to 20 dimensions (c, and d) for the butyrolactone molecule. The normal modes are numbered according to their explained variance (in decreasing order).}
		\label{fig:nm_blac}
	\end{figure}
 
     The second reaction studied is the photoinduced ring-opening of butyrolactone, triggered by excitation to the $S_2$ state. This reaction is characterised by a very rapid ballistic process, with the C-O bond predicted by previous simulations to break within the first 50 fs.\cite{Schalk2020}
 
	\noindent The time evolutions of the electronic populations of $S_0$, $S_1$ and $S_2$ states, as well as the time evolutions of the breaking C-O bond length, which describes the reaction coordinate, for the full- and reduced-dimension dynamics are presented in SI. The compositions of the principal components in terms of normal modes, as well as graphical representations of the first four components calculated from the full-dimensional dynamics are also reported in the SI. The principal components consist of CC stretching, CO stretching, CO out of plane, CH$_2$ rocking and ring torsional motions.

	\noindent Figure \ref{fig:pca_nmv_blac} illustrates the performance of PCA and NMV reduction methods for the Bulac molecule. Insets (a) and (b) in Figure \ref{fig:pca_nmv_blac} display the evolution of the percentage of C-O bonds broken over the course of the dynamics for each dimensional space, for PCA and NMV methods respectively. The quantity of C-O bonds broken is already notably different to the full 30-dimensional dynamics with 24 dimensions selected by NMV (Figure \ref{fig:pca_nmv_blac}(b)), while using PCA reduction notable differences in C-O breakage only appear with fewer than 7 dimensions (Figure \ref{fig:pca_nmv_blac}(a)). PCA is again more accurate than NMV when reducing the dimensionality. The combination of geometric properties Figure \ref{fig:pca_nmv_blac}(d) captures the difference between both methods clearly.

	
	\noindent Looking more closely, using the NMV method, the percentage of C-O bonds broken is no longer well-described decreasing to 22 dimensions, and then becomes significantly worse at 20 dimensions (Figure \ref{fig:pca_nmv_blac}(b)). Not only is the amount of ring-opening decreased, but also some bond reformation occurs (see Figure \ref{fig:pca_nmv_blac}(b) and in SI).
	Figure \ref{fig:nm_blac} shows the 4 (mass weighted) normal modes removed by the NMV method for this decrease from 24 to 20 dimensions.   Interestingly, the first three of these modes seem to exhibit only a small contribution towards the C-O bond breaking process which leads to ring-opening, with larger contributions on the hydrogens and other carbons in the ring. While these three modes do not seem directly involved in the bond breaking, they are probably important in the finalisation of the ring opening and the prevention of ring reclosing. The final mode (d, normal mode 21) is the only one of these four seemingly important modes which contains a significant component on the C-O bond to be broken, and would have been intuitively expected to be necessary to describe the reaction if selecting modes manually.
	
	\noindent This early failure of NMV contrasts to the much better performance of the PCA method, where each PC used as a dimension incorporates weighted contributions of all 30 normal modes. Figure \ref{fig:pca_nmv_blac}(a) shows a slight deterioration of the ring-opening reaction upon removal of principal components 7 and 6. Interestingly, these components contain some CO out-of-plane and torsional motions (see SI).

    \begin{figure*}
		\centering
		\includegraphics[width=\linewidth]{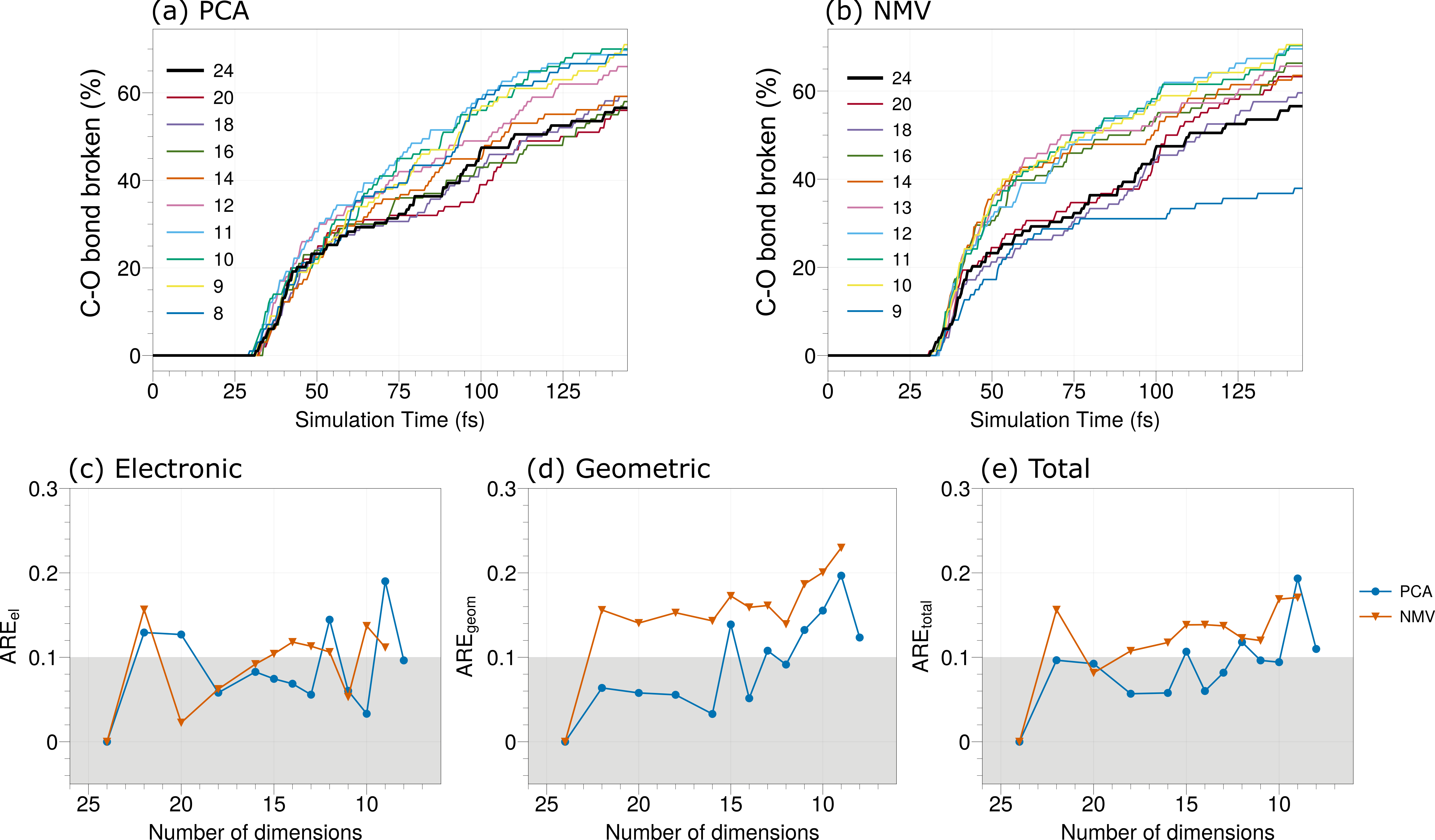}
		\caption{Simulation results for furanone. Evolution (\%) of C-O bond break in function of time (fs) and dimensions for the PCA (a) and the NMV (b) methods. Evolution of the electronic (c), geometric (d) and the total mean (e) errors respectively, as a function of the number of dimensions for both reduction methods. In (c), (d), and (e), the shaded area represents 10\% of relative error compared to the full dimension dynamics.}
		\label{fig:pca_nmv_fur}
	\end{figure*}

    Concerning the electronic properties as seen in Figure \ref{fig:pca_nmv_blac}(c) both methods perform at an essentially equivalent level, with a steady and relatively low (less than 0.2) error throughout. Interestingly, for NMV reduction, the ARE increase to higher values for the geometric properties compared to the electronic ones, in contrast to the results for \textit{trans}-AZM. As a result, the total error as shown in Figure \ref{fig:pca_nmv_blac}(e) shows PCA to be more accurate. From this total error, we can conclude that using PCA we could reduce the full (in NM basis, $3N_{at}-6$) 30-dimensional dynamics to only 7 dimensions, while NMV allows us to reduce to 24 dimensions. It is interesting to note that the error does not necessarily increase monotonically with the decrease of the number of dimensions. We do not believe that this is due to a poor convergence of the dynamics simulations but rather to the alternation of removals of coordinates that enhance and delay electronic relaxation, leading to some kind of cancellation of errors.
	
	\subsection{Ring opening of Furanone (Fur)}
 
    The final reaction studied is the photoinduced ring-opening of the furanone molecule which, like the Bulac, is triggered by excitation to the $S_2$ state. This ring-opening is also characterised by a breaking of a C-O bond. However, unlike Bulac this ring-opening process is more gradual, taking place on a ca.~180 fs timescale due to competition with ring-puckering.\cite{Schalk2020}

    \noindent The time evolutions of the electronic populations of $S_0$, $S_1$ and $S_2$ states, as well as the time evolutions of the breaking C-O bond length, which describes the reaction coordinate, for the full- and reduced-dimension dynamics are presented in SI. The compositions of the principal components in terms of normal modes, as well as graphical representations of the first four components calculated from the full-dimensional dynamics are also reported in the SI. The principal components consist of CO stretching, CH out of plane, and COC bending motions.
    
\begin{figure}
		\centering
        \includegraphics[width=\linewidth]{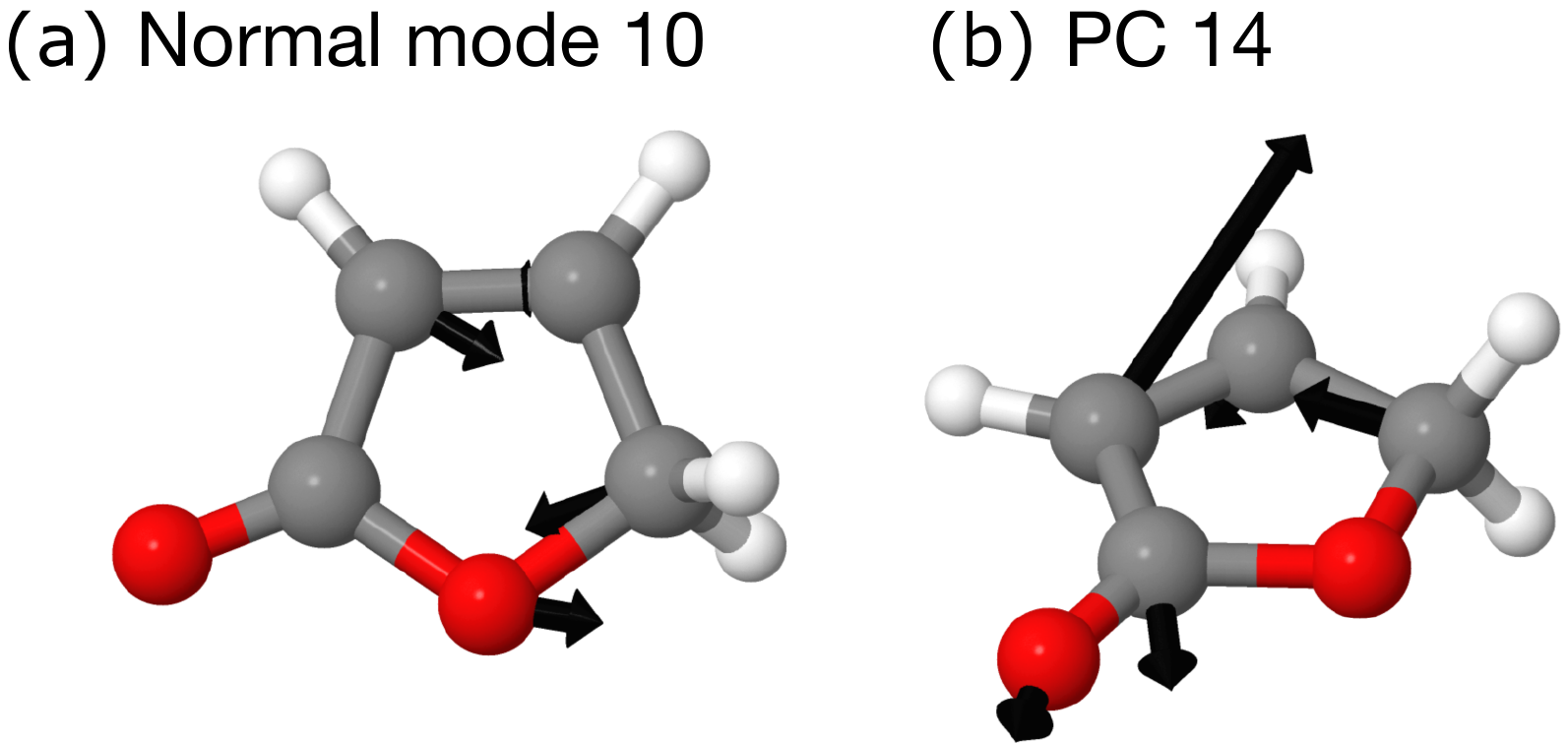}
		\caption{(a) Normal mode 10 removed when reducing from 10 to 9 dimensions using the NMV method, and (b) the PC 14 removed when reducing from 14 to 12 dimensions using the PCA method, for the furanone molecule. The normal modes are numbered according to their explained variance (in decreasing order), consistently with the principal components.}
		\label{fig:nm_fur}
	\end{figure}

    \noindent Figures \ref{fig:pca_nmv_fur}(a) and Figure \ref{fig:pca_nmv_fur}(b) illustrate the evolution of the percentage of C-O bonds broken over simulation time (150 fs), for both the PCA and the NMV methods respectively.
 Using the PCA dimensionality reduction technique, two groups emerge towards the end of the dynamics. The lower percentage group contains the full-dimensionality dynamics, maintaining a similar percentage of C-O breakage by the end of the simulations ($\sim$ 57\%) to as low as 14 dimensions. The higher percentage group is made up of the lower-dimensionality simulations, and is centered around 65\% of C-O bond breakage by the end of the simulations. The most likely explanation for this result is that, when removing additional dimensions below the threshold of 14, part of the competing puckering mechanism is hindered, leading to a faster C-O bond breakage. Figure \ref{fig:nm_fur}(b) depicts one of the PCA modes removed going from 14 to 12 dimensions and seems to support this hypothesis, with the carbons showing out-of-plane motion. Expressed in normal mode coordinates, the principal components 14 and 13 consist of CCC bending, CH out of plane, and CC stretching motions (see SI).
	
 \noindent In contrast, using the NMV dimensionality reduction method, the percentage of C-O bonds broken steadily rises as the dimensionality decreases. This suggests a gradual attenuation of the puckering mechanism with each removed normal mode. However, upon reducing to 9 dimensions, there is a notable decrease in the breakage of C-O bonds (from $\sim$ 60 \%, to $\sim$ 40\%), indicating that the removed normal mode (shown in Figure \ref{fig:nm_fur}(a)) instead principally limits the ring-opening process.
 
	\noindent Upon examining the geometric properties (Figure \ref{fig:pca_nmv_fur}(d)), it is clear that the PCA method allows for a more significant dimensionality reduction, with lower relative error than NMV at all dimensionalities, and accuracy until fewer than 16 dimensions are used. In contrast, the PCA and the NMV methods exhibit quite similar relative errors in electronic properties at all dimensionalities  (Figure \ref{fig:pca_nmv_fur}(c)). The total error, as a result, does indicate that PCA might allow for a reduction in dimensionality, remaining under the 0.1 threshold until reducing to 15 dimensions. However, the similar in scale relative errors in geometric and electronic properties, along with the similar performance for electronic properties, result in the total error giving a less clear indication for possible dimensionality reduction compared to using simply the geometric error. In fact, the quality of the dimension reduction for this reaction seems to be best evaluated by studying simply the percentage of broken C-O bonds. While the results are less clear for this molecule, there seems to be a possible dimensionality reduction from 24 to 16 dimensions using PCA, compared to no clear possible reduction using NMV.

	\section{Discussion}
	\label{sec:disc}

\begin{figure*}
		\centering
		\includegraphics[width=\linewidth]{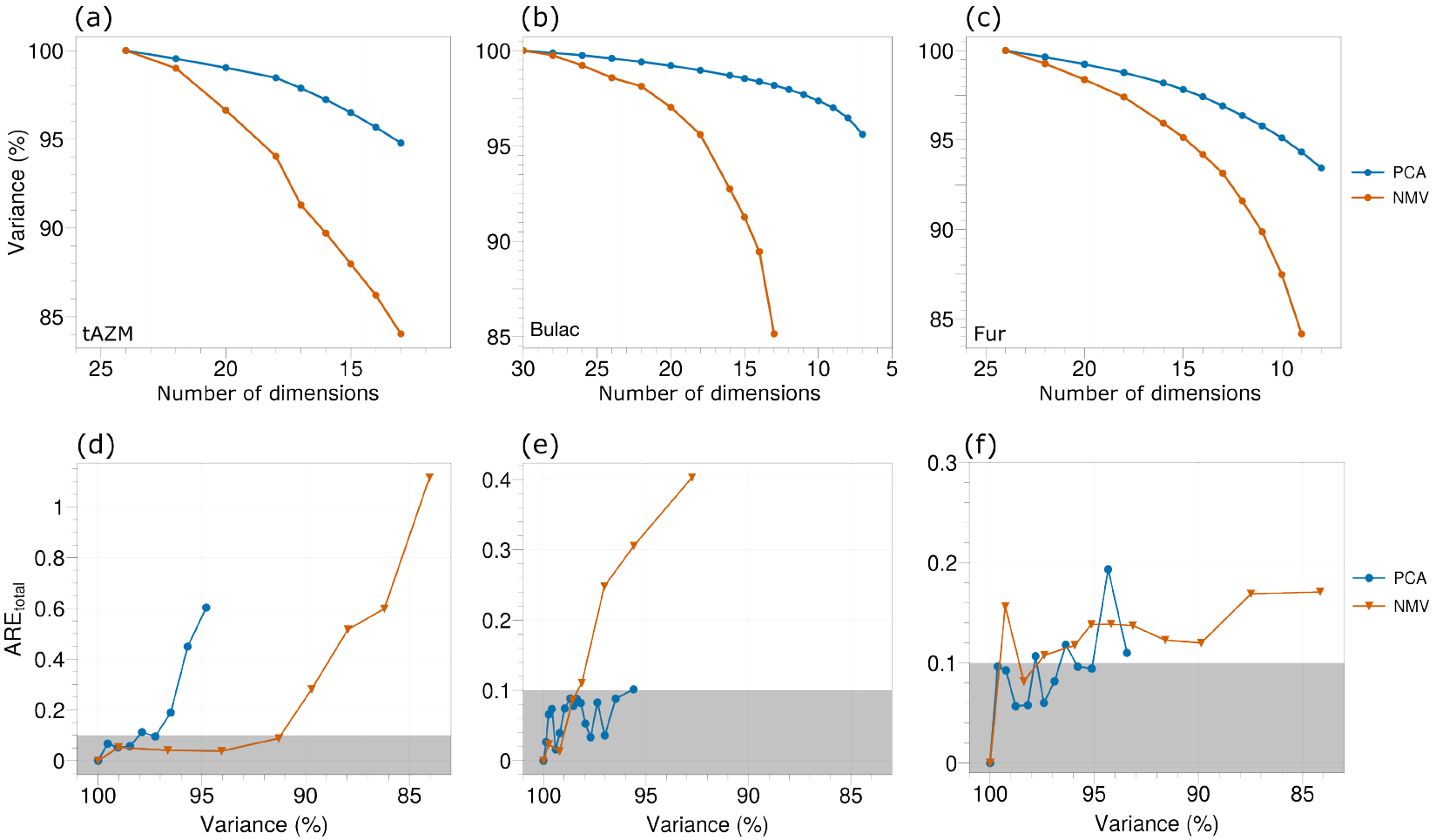}
		\caption{Top panel: evolution of the explained variance (\%) by the PCA (blue) and NMV (orange) methods, as a function of the retained dimensions for (a) trans-azomethane, (b) butyrolactone, and (c) furanone. Bottom panel: evolution of the total error $\text{ARE}_{total}$ with the PCA (blue) and NMV (orange) methods as a function of the explained variance (\%) for (d) trans-azomethane, (e) butyrolactone, and (f) furanone.}
		\label{fig:pca_nmv_all}
	\end{figure*}

    \noindent In this Section, we start by discussing the differences in performances of the two reduction methods. The three molecules studied here all demonstrate that using PCA for dimensionality reduction yields higher quality results compared to the NMV method. This is, of course, expected since PCA is performed in normal mode representation. NMV consists of removing given normal modes while PCA eliminates linear combinations of them, allowing it to lose less variance per removed dimension.
    This is illustrated in Figures \ref{fig:pca_nmv_all}(a-c), which show the quantity of variance retained at each level of dimensionality for NMV and PCA. For instance, in the case of tAZM with only 15 dimensions selected with the NMV method, about 87\% of the variance is kept, whereas with 15 dimensions selected with PCA, over 96\% is still explained. This is also clearly seen for the other reactions in (b) and (c): PCA always leads to a slower loss of variance compared to NMV. 
    
    \noindent However variance is not the whole story, as illustrated in Figures \ref{fig:pca_nmv_all}(d-f) which display the total ARE (\%) as a function of the explained variance (\%) for each molecule and method. For the tAZM molecule (d), the total ARE increases more rapidly with loss of variance under the PCA method compared to NMV, despite the similar accuracy of the two methods as already shown in Section 3.1. In the case of Bulac (e) PCA maintains a total ARE at 10\% down to 95.5\% of kept variance, while NMV crosses the 10\% error threshold at only 98\% variance. For Fur (e), the evolution of error with variance is comparable between the two methods. We do not see a straightforward increase in error as a function of variance which follows a universal trend for both methods - the specific motions removed from the dimension space are more important than the simple value of remaining variance. The variance threshold below which error remains low is dependent on both the method of reduction, and the photochemical reaction, demonstrating the difficulty of establishing a generic variance threshold rule applicable to multiple molecular systems.
	
	\noindent Interestingly, the superior performance of PCA compared to NMV is made clear by different types of property, depending on the molecule. For tAZM, the electronic properties most clearly differentiate between the two methods (Figure \ref{fig:pca_nmv_tazm}(b)), while for Bulac it is clearly the geometric properties which are the most helpful (Figure \ref{fig:pca_nmv_blac}(d)). Finally, for furanone while the differences in geometric properties are clearer than the electronic ones,  the advantage of PCA over NMV is best made obvious from the percentage of C-O broken bonds (Figure \ref{fig:pca_nmv_fur}(a) and (b)). This variability in sensitivity demonstrates an interesting balance between the use of electronic and geometric properties to characterise performance in different reactions, showing a nuanced interplay between molecular and electronic structure depending on the reaction studied. In order to characterise the performance of NMV compared to PCA, properties thought to be representative of the reaction process of interest were selected. This process involves human input. 
	
	\noindent We now discuss more general aspects of the present approach. We note that it relies on an appropriate description of the dynamics by the lower-cost method. We have tested the robustness of the protocol with respect to the number of analysed trajectories in the low-level dynamics in full-dimensionality (see SI for details). Similar ARE were obtained with fewer trajectories, indicating that the original ensemble of trajectories to be analysed does not need to be converged quantitatively. A qualitative description of the process of interest may be sufficient. In principle, any relaxation pathway simulated by the low-level dynamics will be analysed and taken into account in the dimension reduction protocol. However, the presence of a dominating pathway may steer the analysis and thus prevent the occurence of minor pathways upon dimension reduction. If one encounters such a problem, a solution may be to separate the trajectories based on the followed pathway, to perform the dimension analysis separately for the different groups, and finally to merge the extracted subspaces.
	
	\noindent We have also tested the impact of the total simulation time of the original ensemble of trajectories on the accuracy of the dimension reduction (see SI for details). The influence of the simulation time on the total ARE is in general more pronounced when a relatively low number of dimensions is used, whereas it is limited when a large number of dimensions is used. Our recommendation is to set the simulation time such that the whole process of interest is simulated, and to not expect a set of dimensions extracted from short-time dynamics to be relevant for longer time scale.
	
	 In addition, the current protocols are based on normal mode coordinates, since these are expected to perform better than Cartesian coordinates and would allow more direct applications to quantum dynamics simulations. However, normal modes are known to present limitations in case of large amplitude motions. Studying other systems of coordinates would be an interesting topic for a further study.
	 
	  Finally, we also compared the relevant coordinates extracted with the NMV and PCA approaches to gradients and non-adiabatic coupling vectors calculated at the Franck-Condon geometry. The latter quantities are often used to construct ``one shot'' Linear Vibronic Coupling (LVC) models. The results show that the correlation is relatively low: see SI for more details. As previously noted,\cite{Plasser2019} if the dynamics explores regions far away from the Franck-Condon point, these directions are not enough to provide a reliable description of the dynamics of the photoexcited molecule.
	 
	\section{Conclusions}
	\label{sec:concl}
	
	\noindent In this work, we tested the performance of two dimensionality reduction methods, NMV and PCA, in order to identify the most important nuclear coordinates in non-adiabatic processes. The accuracy of both electronic and geometrical properties in reduced-dimensionality dynamics simulations have been assessed on three different photochemical reactions, including both double-bond photoisomerisation and ring-opening processes. In general, both reduction methods allow for the construction, in an automated manner, of lower dimensional spaces which well reproduce the full dimensional dynamics.  
 
    \noindent Using the PCA method, the dimensionality can be reduced from 24 to 16 for tAZM, from 30 to 7 for Bulac, and from 24 to 16 for Fur. Using NMV, the dimensionality can be reduced from 24 to 18 for tAZM, and from 30 to 24 for Bulac, with no clear reduction being possible for Fur. The PCA method always performs better than NMV, with accuracy in dynamics compared to the full-dimensional simulations being preserved when reducing to lower dimensional subspaces. 
	
	\noindent Depending on the molecule, the improvement offered by PCA vs NMV can be more or less significant. For instance, for the tAZM molecule we find that using PCA allows for a reduction of only 1 or 2 additional dimensions compared to NMV, while for the Bulac molecule using PCA allows as many as 15 additional dimensions to be removed compared to NMV, and for the Fur molecule only PCA allows for any removal of dimensions. 
	
	\noindent The use of lower-dimensional spaces optimally constructed through methods such as PCA could provide a viable way to run expensive dynamics simulations, which are often limited due to the curse of dimensionality. One could imagine using the approach outlined in this work (PCA based on TSH variance) to select low-dimensional subspaces built through PCA in which higher cost dynamics could be run. For instance, one could use the selected reduced subspace to simulate the dynamics with more expensive on-the-fly dynamics methods like DD-vMCG or variants of quantum Ehrenfest. Alternatively, one can construct a Hamiltonian model based on the dimension reduction analysis for subsequent MCTDH simulations. We emphasize that building Hamiltonian models remains challenging, and this work only aims to provide guidelines for the first step, the selection of relevant nuclear coordinates.
 In order to generalise our findings further and learn more about which systems lend themselves better to more efficient dimensionality reduction through PCA, further investigations on more diverse and complex molecular systems are required. 

    \section*{Acknowledgements}
    The project is partly funded by the European Union (ERC, 101040356 - ATTOP, M.V., V.D. and A.N.N.). Views and opinions expressed are however those of the authors only and do not necessarily reflect those of the European Union or the European Research Council Executive Agency. Neither the European Union nor the granting authority can be held responsible for them. We also thank the Région des Pays de la Loire who provided post-doctoral funding for A.F., and Nantes University who provided doctoral support for I.C.D.M. The authors thank the CCIPL/Glicid mesocenter installed in Nantes and GENCI- IDRIS (Grant 2021-101353) for the generous allocation of computational time.

	\section*{Supporting Information Available}
	Detailed explanation of both PCA and NMV dimensionality reduction techniques and detailed protocol of dynamics in reduced dimensionality. Raw data for all molecules simulated at all dimensions: time evolutions of electronic populations and key nuclear coordinates, analysis of hopping geometries along key internal coordinates, finer-classication of trans-azomethane conformers, values of each descriptor used for evaluation, averaged over the trajectory ensemble, for each dimensionality in which the dynamics are run for both PCA and NMV reduction methods. Analysis of principal components in terms of normal modes and graphical representations for the first ones. Test of the impact of the simulation time. Test of the robustness of the protocol with respect of the number of analysed trajectories. Comparison with gradient and non-adiabatic coupling vector coordinates.
	
	\bibliography{refs.bib}
\end{document}